**What biology can, and cannot, tell us about conscious AI**

Ulysse Klatzmann[1,2,3] & Adrien Doerig[1,4]
[1] Department of Psychology and Education, Freie Universität Berlin, Berlin, Germany.
[2] Department of Psychiatry and Addiction, Université de Montréal, Montréal, QC, Canada
[3] Azrieli Research Centre, CHU Sainte-Justine, Montréal, QC, Canada
[4] Bernstein Center for Computational Neuroscience, Berlin, Germany

Correspondence: ulysse.klatzmann@umontreal.ca,adrien.doerig@fu-berlin.de

## Abstract

Progress in AI is turning machine consciousness from a philosophical curiosity into a societal issue, and has led to criticism of the widespread computational functionalism framework. Biological Naturalism (BN) claims that biology, not computation, is crucial for consciousness. We discuss which forms of BN are empirically testable. For Type-A-BN, biology "intrinsically" matters for consciousness, without affording unique information processing capabilities. We argue, similarly to the "unfolding argument", that this dissociates consciousness from behaviour, making Type-A-BN untestable. For Type-B-BN, biology matters *because* it affords unique information processing capabilities. Type-B-BN is testable, and not incompatible with computational functionalism. Both face the same task: relating consciousness to information processing. Biology can act as a guide on this quest, but not as a solution.

## Introduction

Artificial Intelligence (AI) is rapidly acquiring sophisticated cognitive abilities, including language, vision, memory, attention, autonomy, and even forms of embodiment. These tantalizing advances have revitalised the age-old question of whether machines can be conscious, turning what was once a theoretical curiosity into a pressing scientific and societal issue.

Within the scientific community, there is broad agreement that current AI systems are not conscious, but far less agreement on why.[1,2]. Importantly, many of the limitations of current systems could in principle be overcome soon[2]. For example, embodiment has been raised as a potentially necessary condition for consciousness[3], which the most advanced AI systems do not currently satisfy. However, there are no in principle barriers to this limitation being overcome in the coming decade(s). To bring the debate beyond necessary conditions that might soon be satisfied, it is essential to reach consensus on a clear scientific framework for determining which systems can be conscious.

Computational functionalism[4] has been the dominant framework in recent decades. On this view, the mind and consciousness can be understood as "software" running on the brain's biological "hardware". Computational functionalism characterizes consciousness in terms of computational organization: consciousness arises when the "right" kinds of computations occur. This framework is well in line with cognitive neuroscience, which studies neural computations under the framework of the brain as an information processing system[5]. However, the advent of AI systems reproducing human computational abilities, arguably passing the Turing test[6] without consciousness[1,2], has shaken the foundations of computational functionalism[1,7–9].

Against this backdrop, Biological Naturalism (BN), often framed as the alternative to computational functionalism, is a different perspective that has recently been gaining traction[1,8,9]. BN shifts the focus from computation to the biological substrate. Consciousness occurs only in the "right" kind of biological systems. While this view comes in many variants, they share the intuition that there is something distinctive about biology that enables consciousness. Hence, according to BN, current AI systems are not conscious, and prevailing AI approaches based on computer chips will never give rise to consciousness.

It is worth mentioning another recent alternative to computational functionalism, based on causal structure[10], which has been criticized for being empirically hard to test[11]. This led to polarized and potentially unsolvable theoretical clashes between paradigms[12,13] instead of evidence-based arguments. To avoid such an outcome, the debate around BN and computational functionalism must be grounded in shared standards of evidence and cumulative science. The key question is therefore not simply whether biology matters, but what empirical predictions follow if it does.

In this spirit, we adopt an empirically oriented methodological stance, asking which forms of BN are empirically testable, and how. We will argue that types of BN claiming that biology is "intrinsically" special cannot be empirically supported. However, types of BN claiming that biology affords unique information processing abilities can. This clarifies the potential role of BN in empirical science as a guide towards which kinds of information processing are crucial for consciousness, rather than as a final answer fundamentally incompatible with computational functionalism.

**Which types of BN can be tested empirically?**

There is no question that many biological features are necessary conditions for *human* consciousness. Nevertheless, if our goal is to answer whether machines can be conscious, we need to ask if they are important for consciousness *in general*[14]. Multiscale integrated processing[8] and astroglia[15] have been suggested to be necessary for conscious processing in the brain. But it remains open whether these features are necessary for consciousness in general. For example, integer addition, like any cognitive function, relies on biological features in the brain[16], but can also be realized without them in calculators.

Like any discussion about consciousness, this debate is inevitably impacted by strong intuitions[17] (Box I). Can it be a coincidence that all the conscious systems we know are biological? How could predicting the next word in a language model lead to consciousness? How could machines be conscious given how different computer chips are from brains? Like all intuitions, these arguments reflect deeply ingrained cognitive heuristics, making them powerful tools for thinking, but also potentially misleading[18]. As discussed in Box I, they do not, by themselves, rule out machine consciousness. Instead, they help identify the questions that an empirical version of BN would need to answer.

There is considerable diversity among views labelled here as BN. Despite this diversity, we can distinguish two types, depending on whether biological systems are proposed to have information processing abilities that other systems do not (see[1] for an alternative classification of BN theories). This dichotomy serves as a wedge between possible variants of BN, and will help clarify which ones can be studied empirically. This distinction is exhaustive, in the sense that any BN proposal necessarily falls into one of them: either biological systems are proposed to have unique information processing abilities, or not. Note also that it is not always clear which type a given BN proposal belongs to, as views are constantly being developed. As such, we do not aim to classify existing BN proposals, but rather to provide a framework to help determine which *types* of BN are empirically testable.

**Type A BN:**

Type A BN proposes that biology does *not* offer any special information processing abilities that non-biological systems don't have. That is, we could in principle replicate the brain's information processing in machines. Nevertheless, only the biological implementation would be conscious. According to Type A BN, what matters is what biology is made of, not what it does.

One version of this view is based on the idea that two systems can be equivalent computationally but differ subcomputationally[9]. For example, an AND gate can be implemented in an electrical circuit, or using water pipes. Both implement the same information processing, but differ subcomputationally. This version of type A BN would then propose that biological subcomputational features are what matters for consciousness, instead of features of information processing.

We argue that this claim cannot, even in principle, be tested empirically, for reasons related to the "unfolding argument"[11] (see also [19]). Indeed, empirical tests of consciousness all ultimately rely on behavioural reports[17,19–25]. When studying the contents of consciousness, we present stimuli and ask observers to report what they perceive. When studying states of consciousness, we use behavioural indicators, such as the bispectral index to quantify coma depth. Yet Type A BN grants that biological and non-biological systems could, in principle, implement the same information processing, leading to identical behaviour, and therefore identical experimental results about consciousness, while still maintaining that only biological systems will be conscious.

To illustrate the problem, imagine a scenario in which part of a patient's brain that is crucial for consciousness is replaced by a digital computer chip. If the chip performs equivalent information processing (which is possible by hypothesis in Type A BN), the brain as a whole operates as usual, leading to identical reports about consciousness. Yet, Type A BN claims that consciousness has disappeared or has been altered. The problem is that this claim cannot be supported by experimental evidence, since nothing observable changed in the reports of the subject about their consciousness. The patient still insists on being conscious, still behaves identically in all experiments, etc. But Type A BN claims that none of this matters, because biology is absent - an empirically untestable claim (see[11,21,25–28] for related discussions).

As far-fetched as this example sounds, we are getting closer to real world cases. Researchers recently implanted electrodes in the occipital cortex of a blind patient, and successfully evoked simple conscious visual percepts[29] (more detailed percepts were evoked in monkeys[30]). How do we know that conscious percepts were evoked? Because the patient reported experiencing them. But Type A BN implies the possibility that, because the relevant biological aspects might not have been in place, the patient might merely have reported these percepts without experiencing them consciously. The problem is that, under Type A BN, we can never know if this is the case. The Type A BN proponent must always doubt such reports on the grounds of the non-biological nature of the implant, no matter how reliably the patient claims to be conscious.

While this is still a rudimentary case, more powerful implants are increasingly plausible, and the field of brain machine interfaces is making great strides[31]. In the not so distant future, more important aspects of brain information processing may be possible to offload to artificial systems. This could have serious impacts on consciousness, according to Type A BN, that can never be detected empirically, even in principle.

In summary, Type A BN implies that consciousness cannot be detected through empirical measurements, but should be inferred based on the biological substrate. In principle, this could be true, but we can never gather any scientific evidence to support it, making it scientifically inert (see also[21,24] for related arguments, and the debate around overflowing phenomenal consciousness[32]). This precludes empirical guardrails, likely leading to unresolvable fragmentation, endless debates and ultimately to a loss of credibility for the field[33,34].

**Type B BN:**

Type B BN proposes that biology *does* offer unique information processing abilities. That is, we could not implement what the brain does in non-biological systems, even in principle, because there are key differences between biological and non-biological systems[16]. According to Type B BN, what matters is what biology does, not what it is made of. For this reason, Type B BN proposes, only biological systems can be conscious. Two questions arise: (1) Are these proposed differences between biological and non-biological systems real? (2) Are there reasons to believe that these differences matter for consciousness? Here, we set question (1) aside and accept for the sake of argument that biological systems have unique abilities, in order to focus on their importance for consciousness.

To make this point more concrete, we consider one frequently cited difference between biological and artificial systems: digital computers operate on discrete bits of information, while biological systems exhibit continuous analog dynamics, over time (neurons can spike at any moment), space (membrane potentials and chemical gradients vary continuously), and magnitudes (membrane potentials can be any continuous number)[1,9]. This difference may matter in chaotic systems, where even infinitesimally small differences in initial conditions can lead to large divergences over time[35]. Hence, the continuous nature of biological dynamics, compared with the discrete nature of digital computers, has been proposed as a relevant difference by proponents of BN: because digital systems represent states with finite precision, they cannot reproduce exactly the trajectory of a continuous chaotic system, and small discrepancies may lead to widely different outcomes[1].

Crucially, however, the existence of a difference between biological systems and computers does not entail that this difference matters for consciousness. Chaotic systems do not generate arbitrary behaviour. Their dynamics are constrained by attractors that determine the trajectories the system can follow. Digital systems can therefore diverge from continuous counterparts while still preserving the long-time statistical behaviour. Climate science provides a useful analogy. Weather systems are chaotic, meaning that even tiny numerical differences quickly lead to vastly different weather predictions. Hence, we all agree that the analog nature of the atmosphere is different from the discrete nature of computers. However, despite these differences, computers can capture important features of weather systems, and this is precisely what makes weather and climate modelling scientifically useful. We explicitly set aside the different debate on whether a simulated hurricane is "wet"; what matters is that discrete computers can capture important features of hurricanes, such as their size, speed, formation dynamics, or the eye of the storm[17,36–38]. Likewise, despite real differences between the analog nature of the brain and the discrete nature of computers, the latter may still be able to capture what matters for consciousness. Alternatively, they might not be able to capture some crucial features of consciousness, but arguments and experiments are needed to support this claim (see next section).

As another example, autopoiesis has been argued to be a defining property of living beings that is absent in current AI systems[1]. Assuming that autopoiesis is a uniquely biological property, and that it is necessary to enable specific information processing capacities, for example embodied regulation, or organism-level integration, this is a valid difference between biological and non-biological systems. But the existence of this difference does not show that it matters for consciousness. To become empirically testable, the proposal must specify which capacities autopoiesis enables, and how these capacities should affect conscious processing. Alternatively, if autopoiesis is treated as sufficient for consciousness independently of any observable consequences, the proposal collapses into Type A BN and faces the corresponding challenges.

This argument generalises to any other proposed difference between biological and non-biological systems. Even accepting that there are differences, it is not enough to just point out that they exist. We need to empirically show that they matter for consciousness.

Importantly, a type B BN proposal needs to answer both questions (1) and (2) above - the feature proposed to be crucial for consciousness needs to be (1) unique to biology and (2) relevant for consciousness. Several recent proposals take steps in this direction. For example, dendritic integration has been experimentally linked to consciousness[39] (addressing question 2) - but it has not been shown that the features of dendritic integration that are crucial for consciousness cannot be implemented in non-biological systems (leaving question 1 open). Conversely, there are arguments proposing that certain aspects of biology may be extremely difficult to implement in machines[8,16] (addressing question 1), but it has not been shown that these very same uniquely biological features matter for consciousness (leaving question 2 open).

In summary, we argue that Type A BN does not generate empirically testable predictions about consciousness. In contrast, Type B BN can be safely grounded in empirical research, provided that proponents specify which differences between biological and non-biological systems matter for consciousness, and test these claims empirically. In the absence of empirical evidence directly testing the involvement of uniquely biological features in consciousness, BN remains a theoretical possibility, but is not a supported theory.

**Current evidence**

What would empirical tests of BN look like? In consciousness science, empirical evidence comes from contrasting conscious vs. non-conscious processing[20,25,40]. This applies to both the state (e.g. dreamless sleep vs. dreaming) and content (e.g. seen vs. unseen targets in psychophysical experiments) of consciousness. Theorists need to show that, when a variable proposed to matter for consciousness changes, then consciousness changes as predicted. For example, Global Workspace Theory[41] can predict "when there is no ignition, the target is unseen, when there is ignition, the target is seen"[42]. While there remain important debates about such experiments[43], they illustrate how a theory can gain empirical traction by linking a proposed mechanism to specific contrasts between conscious and non-conscious processing.

As discussed, several theoretical candidate differences between biological and non-biological systems have been suggested, from dendritic integration[44] to multiscale brain dynamics[8]. To gather empirical support, BN needs to pinpoint aspects unique to biology *and* show that they are crucial for consciousness. They should predict "when this biological feature X is absent there is no consciousness, when it is present, there is consciousness". Going back to the example of continuous vs. discrete computations discussed in the previous section, BN needs to say why this matters. Is it because of chaos? If so, why is chaos crucial for consciousness, given that it is not crucial, for example, for the formation of hurricanes? One possibility would be to predict that when the system is at the wrong location of its chaotic attractor, there is no consciousness, when it is at the right location, there is consciousness. Then, this prediction needs to be tested. Analogous reasoning applies to other proposed relevant differences between biological and non-biological systems.

**Biology plays an enabling role, not a fundamental one, in testable BN theories**

Type A and Type B BN take fundamentally different views on how to determine whether only biological systems can be conscious. Type A BN proposes to infer consciousness from the biological substrate, whereas Type B BN asks which biological properties impact information processing relevant to consciousness.

The above considerations have two conclusions: (1) BN can be safely grounded in empirical science, but (2) only if the relevant biological properties have an impact on information processing. Hence, BN cannot support the claim that biology is "intrinsically" special. What matters is what the biology does (Type B), not what it is made of (Type A). What matters is not the substrate *qua* substrate, but that it may enable information processing abilities that are impossible or inefficient in current computers. This imposes clear boundaries on how to interpret the BN research program. We will never be able to determine if a system is conscious based solely on whether or not it is biological. The central challenge is still to understand which kinds of information processing matter for consciousness. BN can act as a guide on this quest, but not as an answer.

These arguments do not reject contemporary BN proposals. Indeed, many adopt a Type B view, by treating biological features as candidate mechanisms that enable specific kinds of processing crucial for consciousness, rather than as intrinsically consciousness-making substances[8,16]. Our contribution is threefold: (1) we propose that the Type A/B distinction is essential to understanding which types of BN are testable, (B) we provide an argument against the empirical testability of Type A BN, (C) we explicitly state a methodological requirement for Type B BN: such proposals become scientific hypotheses only when they specify which biological capacities matter, how they affect conscious processing, and how these claims could be tested.

This reframes the traditional opposition between BN and computational functionalism. Arguments against computational functionalism are often taken to count as arguments in favour of BN[1,9]. However, Type B BN and computational functionalism are not necessarily mutually exclusive[8]. Both could be true, if the forms of information processing required for consciousness can only be implemented by biological systems. This Type B BN scenario is compatible with (certain forms of) biologically-constrained computational functionalism[8]. Equally, both could also be false: consciousness might depend on properties that are not captured by abstract computation, but are not exclusive to living systems either. Hence, arguments against computational functionalism should not automatically be taken to support BN, and vice-versa.

## Conclusion

The rapid rise of AI raises the question of artificial consciousness, and poses huge ethical challenges which we should address before it is too late. A common argument for BN is that we should avoid false positives—wrongly attributing consciousness to machines[1]. This concern is legitimate, but too readily attributing consciousness to biological systems such as embryos raises equally challenging ethical issues. Hence, consciousness should not be given to machines “for free”, but it should not be taken as a basic biological property without specifying the relevant mechanisms either.

To progress on these questions, we need to understand which differences between brains and current AI are crucial for consciousness. This is a deep challenge, as current AI systems differ from humans in endless ways. Biological implementation is indeed one of them, but, from a computational perspective, there are enormous differences too. Hence, there are other options than BN to show differences between biological and AI systems. BN should be held to the same high standards of empirical testing as computational proposals. Currently, there is arguably little empirical evidence for a crucial role of uniquely biological features for consciousness. Deriving paradigms to test this should be a priority of BN.

Importantly, empirically testable Type B BN is not necessarily incompatible with computational functionalism. Hence, we should not spiral into an entrenched clash of paradigms based on theory and intuitions. BN draws our attention to biological processes while computational functionalism focuses on the computer metaphor of the brain. But these views do not offer any easy ultimate solutions to the problem of consciousness. We still all face the same daunting task: understanding which kinds of information processing matter for consciousness.

We are living through strange and puzzling times, and we should be open to new ideas. But crucially, consciousness and AI researchers can better solve these questions if they work together. We should not isolate consciousness science from the rest of science. We should not entangle ourselves into a veil of biological exceptionalism. As the eyes of the world turn to the science of consciousness for answers, we need to avoid dogmatism and try to reach consensus. To this end, it is crucial to ground debates in empirical evidence.

**Box I: Intuitive arguments are insufficient to establish the impossibility of artificial consciousness without empirical testing**

*Intuition 1: Induction from known biological cases*

All conscious systems we currently know are biological. From this, it may be tempting to infer that biology is crucial. However, showing that all known conscious systems share property X does not establish that X is crucial for consciousness. For example, all conscious systems we know of live on earth, but that is not an argument showing that only the earth can support consciousness. Hence, which biological features (if any) are crucial for consciousness cannot be settled by induction from biology alone, but must be argued for and tested empirically.

*Intuition 2: The re-description fallacy*

AI systems are sometimes deemed incapable of consciousness because they are "just" doing something simple. For example, "just trained to predict the next word". By the same logic, humans could be described as "just being selected to pass on their genes". As Millière and Buckner[45] put it:

"*[The re-description] fallacy arises when critics argue that a system cannot model a particular cognitive capacity [$\phi$], simply because its operations can be explained in less abstract and more deflationary terms. [...] Such arguments are only valid if accompanied by evidence demonstrating that a system, defined in these terms, is inherently incapable of implementing $\phi$. To illustrate, consider the flawed logic in asserting [...] that brain activity could not possibly implement cognition because it can be described as a collection of neural firings.*"

Hence, re-descriptions are insufficient on their own to deny artificial consciousness. Arguments and empirical data are needed to show which consciousness-relevant features (if any) are impossible for artificial systems.

*Intuition 3: From difference to exclusivity*

There are real differences between biological and artificial systems. For example, computers are based on discrete computation, which differs from the continuous, analog nature of biology. BN proponents claim that these differences are important for consciousness. Here, one must be wary of the motte-and-bailey fallacy, which is a rhetorical move where one advances a strong claim (the bailey), but when challenged retreats to a weaker claim (the motte). Thus, the strong claim is never actually defended. In the case of BN, one cannot advance a strong BN claim (only biological systems can be conscious), while in practice only defending a weaker claim (there are differences between artificial and biological systems). There is a huge gap between these claims, since differences between biological and non-biological systems may not matter for consciousness. Arguments and empirical data are needed to show which differences (if any) matter.

**End of Box I**

**Acknowledgments**

U.K. was supported by a START Grant from Freie Universität Berlin ("Cortical Self Organization in Neural Network Models") and by an Impact+ Canada Postdoctoral Research Training Award from the Natural Sciences and Engineering Research Council of Canada (NSERC/CRSNG; application no. BFRIC-P-614238-2026), hosted at CHU Sainte-Justine / Université de Montréal. We also thank MESEC (the Mediterranean Society for Consciousness Science) for fostering discussions and reflections that helped shape this article.